\documentclass[twocolumn,aps,superscriptaddress,nofootinbib,floatfix]{revtex4}
\usepackage{epsfig,bm,feynmf}
\usepackage{graphics}
\usepackage{amsmath}
\usepackage{mathrsfs}
\usepackage{appendix}
\usepackage{bm}
\usepackage[normalem]{ulem}  % \sout{old text} for strikeout
\usepackage[dvips]{color} % For blue in-text comments and additions

\renewcommand{\sout}{\bgroup \color{red} \ULdepth=-.5ex \ULset}

\begin{document}
\title{Anomalous transport model study of chiral magnetic effects in heavy ion collisions}	

% authors
\author{Yifeng Sun}
\email{sunyfphy@physics.tamu.edu}
\affiliation{Cyclotron Institute and Department of Physics and Astronomy, Texas A$\&$M University, College Station, Texas 77843, USA}%

\author{Che Ming Ko}
\email{ko@comp.tamu.edu}
\affiliation{Cyclotron Institute and Department of Physics and Astronomy, Texas A$\&$M University, College Station, Texas 77843, USA}%

\author{Feng Li}
\email{fengli@comp.tamu.edu}
\affiliation{Cyclotron Institute and Department of Physics and Astronomy, Texas A$\&$M University, College Station, Texas 77843, USA}%

% date
\date{\today}% It is always \today, today,

\begin{abstract}
Using an anomalous transport model for massless quarks and antiquarks, we study the effect of a magnetic field on the elliptic flows of quarks and antiquarks in relativistic heavy ion collisions. With initial conditions from a blast wave model and assuming that the strong magnetic field produced in non-central heavy ion collisions can last for a sufficiently long time, we obtain an appreciable electric quadrupole moment in the transverse plane of a heavy ion collision.
The electric quadrupole moment subsequently leads to a splitting between the elliptic flows of quarks and antiquarks.  The slope of the charge asymmetry dependence of the elliptic flow difference between positively and negatively charged particles is positive, which is expected from the chiral magnetic wave formed in the produced QGP and observed in experiments at the BNL Relativistic Heavy Ion Collider, only if the Lorentz force acting on the charged particles is neglected and the quark-antiquark scattering is assumed to be dominated by the chirality changing channel.
\end{abstract}

\maketitle

\section{introduction}

Experiments at the BNL Relativistic Heavy Ion Collider (RHIC)~\cite{Adams:2005dq,Adcox:2004mh,Arsene:2004fa,Back:2004je} and the CERN Large Hadron Collider (LHC)~\cite{Aamodt:2008zz} have provided convincing evidence for the formation of a quark-gluon plasma (QGP) during the early stage of a heavy ion collision.  Extensive studies on the anisotropic flows of various particles~\cite{Heinz:2013th} and the nuclear modification factor of energetic jets~\cite{Jacobs:2004qv} have shown that the produced QGP, which has near zero baryon chemical potential, behaves almost like an ideal fluid with a viscosity to entropy density ratio not much larger than the theoretical lower bound derived from strongly interacting quantum field theories~\cite{Kovtun:2004de}. Because of the restoration of chiral symmetry in QGP and the strong magnetic field produced in noncentral heavy ion collisions, almost massless light quarks can lead to the chiral separation effect (CSE) in which axial charges are separated along the direction of the magnetic field if the electric charge chemical potential is nonzero~\cite{PhysRevD.70.074018,PhysRevD.72.045011,PhysRevLett.103.191601}.  The massless quarks can also lead to the chiral magnetic effect (CME) in which electric charges are separated along the direction of magnetic field if the axial charge chemical potential is nonzero~\cite{Kharzeev2008227,PhysRevD.78.074033,Kharzeev2010205},  Through the interplay between the CME and the CSE, a new form of gapless collective excitation, called the chiral magnetic wave (CMW), emerges from the coupling of the density waves for axial and electric charges~\cite{PhysRevD.83.085007,PhysRevLett.107.052303}.

In a heavy ion collision, the chiral magnetic wave can induce a charge quadrupole moment in the transverse plane perpendicular to the beam direction and lead to the splitting of the elliptic flows of positively and negatively charged particles during the expansion of the produced QGP~\cite{PhysRevLett.107.052303}. By solving the wave equation for the charge density in a schematic model for heavy ion collisions, a splitting of elliptic flows has been found with a magnitude comparable to that measured in experiments if the magnetic field is sufficiently strong and lasts long enough~\cite{PhysRevLett.107.052303}.  For a more quantitative study, the anomalous hydrodynamics, which extends the normal hydrodynamics to include also the dynamics of chiral magnetic wave, has been developed~\cite{PhysRevLett.103.191601} and applied to study its  effect on elliptic flows in heavy ion collisions using initial conditions from a blast wave model and assuming a strong and long-lived external magnetic field~\cite{Hongo:2013cqa,PhysRevC.89.044909}.  However, contradicting conclusions about the splitting between the elliptic flows of positively and negatively charged particles have been obtained with Ref.~\cite{PhysRevC.89.044909} requiring the chiral magnetic effect and Ref.~\cite{Hongo:2013cqa} without requiring such an effect.

Also, microscopic chiral kinetic equations have been developed~\cite{PhysRevLett.109.162001,Son:2012wh,Son:2012zy,PhysRevLett.109.232301,PhysRevLett.110.262301,Manuel:2014dza}, which can include the nonequilibrium effect on the chiral magnetic wave in heavy ion collisions.  In Ref.~\cite{PhysRevLett.109.162001}, the chiral kinetic equation is derived from considering the Berry phase associated with the action of a quantum system in an external magnetic field. The resulting chiral kinetic equation reproduces the chiral anomaly as well as the CME and CSE in equilibrium systems. However, this equation is not manifestly Lorentz covariant and requires the introduction of a modified Lorentz transformation~\cite{PhysRevLett.113.182302}.  The same chiral kinetic equation can be derived from the Landau fermi liquid theory by taking into account the Berry curvature flux through the fermi surface~\cite{Son:2012wh,Son:2012zy} or from the semiclassical Foldy-Wouthuysen diagonalization of the quantum Dirac Hamiltonian for massless fermions and antifermions~\cite{Manuel:2014dza}. On the other hand, a manifestly Lorentz covariant chiral kinetic equation has been derived in Refs.~\cite{PhysRevLett.109.232301,PhysRevLett.110.262301} based on the consideration of the covariant Wigner function for massless spin 1/2 fermions, which reproduces the non-covariant chiral kinetic equation after integrating over the zeroth component of the four-momentum.

In the present study, we solve the noncovariant chiral kinetic equation for massless quarks and antiquarks using the test particle method~\cite{Wong:1982prc} and include also the scatterings of these particles by assuming that scatterings are not affected by the magnetic field.  With initial conditions taken from a blast wave model, we then use this anomalous transport model to study the elliptic flow in heavy ion collisions at the highest energy from RHIC and compare the results with those obtained from the anomalous hydrodynamics~\cite{PhysRevC.89.044909} and with the experimental data~\cite{Wang2013248c}.

This paper is organized as follows. In the next section, we show that under some simple assumptions, the equations of motion for the test particles obtained from the chiral kinetic equation can be derived from considering the rate of change of the total angular momentum of a test particle, which includes both its orbital and spin angular momenta. We then describe in Sec. III the details on how heavy ion collisions are modeled in the present study. In Sec. IV, results from our study are presented and compared with those from the anomalous hydrodynamics and the experimental data.  We also discuss  in this section the effects due to the Lorentz force and the chirality changing scattering.  Finally, a summary is given in Sec. V.

\section{The chiral kinetic equation}

The classical equations of motion for a massless spin-1/2 particle in an external electromagnetic field were firstly derived in Ref.~\cite{PhysRevLett.109.162001} using the path-integral formulation. In the adiabatic approximation, this leads to a vector potential or Berry connection ${\bf A}_p$ in  momentum space~\cite{Berry:1984jv}. The corresponding Berry curvature ${\bf b}=\nabla_p\times{\bf A}_p$ then modifies the velocity of a massless spin half particle to
\begin{eqnarray}
&&\frac{d\mathbf{r}}{dt}=\hat{\mathbf{p}}+\dot{\mathbf{p}}\times\mathbf{b},
\label{1}
\end{eqnarray}
where $\hat{\bf p}$ is a unit vector along the direction of the momentum and ${\bf b}=\pm\frac{\hat{\mathbf{p}}}{2p^2}$ with the plus and minus signs for particles with spin parallel (positive helicity) and antiparallel (negative helicity) to the momentum, respectively. We have used in the above the convention $\hbar=c=1$.

Equation~(\ref{1}) can also be derived from the change of the total angular momentum of a massless spin-1/2 particle due to an external force. Since the total angular momentum of such a particle of positive or negative helicity is ${\bf J}={\bf L}+{\bf S}={\bf r}\times{\bf p}\pm\hat{\bf p}/2$, its rate of change is
\begin{eqnarray}
\frac{d\mathbf{J}}{dt}&=&\frac{d\left(\mathbf{r}\times\mathbf{p}\pm\frac{\hat{\mathbf{p}}}{2}\right)}{dt}\nonumber\\
&=&\frac{d\mathbf{r}}{dt}\times\mathbf{p}+\mathbf{r}\times\frac{d\mathbf{p}}{dt}\pm\left[\frac{\dot{\mathbf{p}}}{2p}
-\mathbf{p}\left(\frac{\mathbf{p}}{2p^3}\cdot\dot{\mathbf{p}}\right)\right]\nonumber\\
&=&\left(\dot{\mathbf{r}}\mp\dot{\mathbf{p}}\times{\frac{\mathbf{p}}{2p^3}}\right)\times{\mathbf{p}}+\mathbf{r}\times\mathbf{F},
\end{eqnarray}
where we have used the relation $d{\bf p}/dt={\bf F}$ with ${\bf F}$ being the external force. Using the fact that $d{\bf J}/{dt}={\bf r}\times{\bf F}$, we then have $\dot{\mathbf{r}}\mp\dot{\mathbf{p}}\times\frac{\mathbf{p}}{2p^3}=f(p)\hat{\bf p}$, where $f(p)$ is a function of the magnitude $p$ of the momentum.  In the absence of external force, ${\bf F}=0$, one has $\dot{\bf r}=\hat{\bf p}$, which then requires $f(p)=1$. This thus leads to $\dot{\mathbf{r}}=\hat{\bf p}\pm\dot{\mathbf{p}}\times\frac{\mathbf{p}}{2p^3}=\hat{\bf p}+\dot{\bf p}\times{\bf b}$,
which is the same as Eq.~(\ref{1}).

Including the Lorentz force due to an external magnetic field ${\bf B}$, ${\bf F}=Q\dot{\bf r}\times{\bf B}=\dot{\bf p}$, with $Q$ being the charge  of the massless spin-1/2 particle, its equations of motion can then be obtained from Eq.~(\ref{1}), and they are given by
\begin{eqnarray}\label{chiral}
\frac{d\mathbf{r}}{dt}&=&\frac{\hat{\mathbf{p}}+Q(\hat{\mathbf{p}}\cdot\mathbf{b})\mathbf{B}}{1+Q\mathbf{B}\cdot\mathbf{b}},\\
\frac{d\mathbf{p}}{dt}&=&\frac{Q\hat{\mathbf{p}}\times\mathbf{B}}{1+Q\mathbf{B}\cdot\mathbf{b}}.
\end{eqnarray}
These equations are the same as those given in Refs.~\cite{PhysRevLett.109.162001,PhysRevLett.109.232301,PhysRevLett.110.262301,Manuel:2014dza}.

\begin{figure}[h]
\centering
\includegraphics[width=0.4\textwidth]{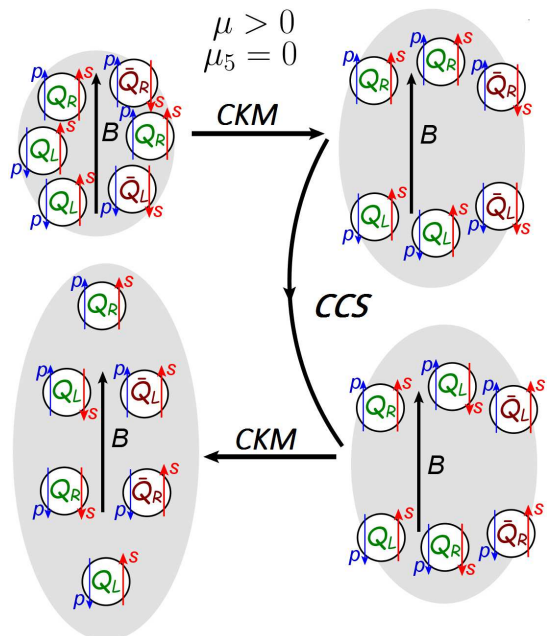}
\caption{(Color online) Effects of chiral kinetic motion (CKM) and chirality changing scattering between quark and antiquark (CCS) on the distribution of massless quarks and antiquarks in the transverse plane of a heavy ion collision with initial positive charge chemical potential $\mu>0$ and vanishing axial charge chemical potential $\mu_5=0$.}
\label{cme}
\end{figure}

To apply the chiral kinetic equation to heavy ion collisions, we also need to include the scattering of massless quarks and antiquarks, which has so far only been addressed in the relaxation time approach either for quark-quark scattering~\cite{Satow:2014lva} or for quark-gluon scattering~\cite{Manuel:2015zpa}.  Since the interaction Lagrangian of massless quarks and antiquarks with gluons in QCD is ${\cal L}=ig \bar q_R{A\mkern-8mu/}q_R+ig \bar q_L{A\mkern-8mu/}q_L$, where $g$ is the strong coupling constant, $q_R$ and $q_L$ are the field operators for quarks of right and left chiralities, respectively, and ${A\mkern-8mu/}=\gamma^\mu A_\mu$ with $A_\mu$ being the gluon filed, there are no changes in the chiralities of quarks in quark-quark scattering and of antiquarks in antiquark-antiquark scattering. For the scattering between quark and antiquark, their chiralities can, however, change through the $s$-channel annihilation process.  This process ensures both local charge and axial charge conservations in an equilibrated quark matter and is essential for generating a splitting of the elliptic flows of positively and negatively charged particles in the kinetic approach.  The reason for this is as follows.  Although the chiral kinetic motion can lead to a separation of particles of right chiralities from those of left chiralities in space, the effect is the same for positively and negatively charged particles. Only by including quark-antiquark scatterings that change their chiralities can result in a charge quadrupole moment in the transverse plane of a heavy ion collision. For simplicity, we thus only allow massless quarks and antiquarks to undergo chirality changing scattering in quark-antiquark scattering.

The combined effects due to the chiral kinetic motion (CKM) and the chirality changing scattering (CCS) between quark and antiquark in heavy ion collisions in the presence of a magnetic field are illustrated in Fig.~\ref{cme}. The upper left picture shows the initial distribution of quarks and antiquarks in the transverse plane for the case of positive charge chemical potential $\mu>0$ and vanishing axial charge chemical potential $\mu_5=0$, corresponding to more quarks than antiquarks but with equal number of right and left chiralities.  In the above, we have $\mu=(\mu_R+\mu_L)/2$ and $\mu_5=(\mu_R-\mu_L)/2$, where $\mu_R/T=(N_R-N_{\bar R})/(N_R+N_{\bar R})$ and $\mu_L/T=(N_L-N_{\bar L})/(N_L+N_{\bar L})$ with $N_R$ and $N_L$ denoting the numbers of particles with right and left chiralities, respectively, while $N_{\bar R}$ and $N_{\bar L}$ denoting corresponding numbers for antiparticles.  The temperature of the quark matter is denoted by $T$.  Equation (3) then indicates that quarks and antiquarks of right chiralities would move upward along the direction of the magnetic field, while those of left chiralities would move in the opposite direction, leading to a positive axial charge dipole moment in the transverse plane, as shown in the upper right picture.  The lower right picture shows that the effect of the chirality changing quark-antiquark scattering $q_R\bar q_R\to q_L\bar q_L$ and $q_L\bar q_L\to q_R\bar q_R$.  Further upward and downward motions of quarks and antiquarks of right and left chiralities, respectively, according to Eq. (3) result in the lower left picture, which clearly shows a positive axial charge dipole moment and a positive charge quadrupole moment in the transverse plane.  The latter can then lead to a splitting of the elliptic flows of positively charged quarks and negatively charged antiquarks as shown in Refs.~\cite{PhysRevLett.107.052303,Ma:2014iva}.

In the above discussion, we have not considered the effect of gluons in the system. Since gluons have spin one, they will not change the chiralities of the quarks or antiquarks that they scatter with~\cite{Manuel:2015zpa}. Including these scatterings will thus not affect the picture discussed in the above. Also,
the magnetic field affects the scattering of quarks and antiquarks through the change of the phase-space integral in the collision term of the chiral transport model~\cite{Satow:2014lva,Manuel:2015zpa}, i.e., replacing the usual $d^3{\bf x}d^3{\bf p}/(2\pi)^3$ by $(1+Q{\bf B\cdot{\bf b}})d^3{\bf x}d^3{\bf p}/(2\pi)^3$, which is needed to ensure the correct equilibrium distribution in the presence of the magnetic field.  Since it is not yet known how this effects can be included in the chiral transport model, we thus neglect them in the present study.

To implement particle scatterings in the chiral or anomalous transport model, we generalize the geometric method of Ref.~\cite{Bertsch:1988ik} by using the scattering cross section $\sigma$ in the fireball frame to check whether the impact parameter between two colliding particles is smaller than $\sqrt{\sigma/\pi}$ and if the two colliding particles pass through each other at the next time step during the evolution of the system.   For the three-momenta of the two particles after their scattering, they are taken to be isotropic in their center-of-mass frame.  We neglect, however, the Pauli blocking effect on the final states because of the high temperature of the partonic matter produced in heavy ion collisions.

\section{Heavy ion collisions in the presence of a magnetic field}

To apply the anomalous transport model to heavy ion collisions, we use a blast wave model for the initial conditions. Specifically, we take the cubic power of temperature distribution, which is proportional to the particle number density distribution, in the transverse plane of the collision to have a Woods-Saxon form
\begin{eqnarray}
T(x,y)=\frac{T_0}{(1+e^{\frac{\sqrt{x^2+y^2/c^2}-R}{a}})^{\frac{1}{3}}},
\end{eqnarray}
where $c$ describes the spatial anisotropy of produced partonic matter in the transverse plane of non-central heavy ion collisions, $R$ is the radius, and  $a$ is surface thickness.  The transverse momentum distributions of quarks and antiquarks are then given by the corresponding Boltzmann distributions with a charge chemical potential $\mu$.  For simplicity, we take quarks and antiquarks to have same electric charge of $e/2$ and $-e/2$\footnote{Although this would underestimate the effect on $u$ and $\bar u$ quarks and overestimate that on $d$ and $\bar d$ quarks, the net effect would be minimal for flavor symmetric quark matter.}, respectively, and assume that the ratio $\mu/T$ is uniform in space.  For distributions in the $z$ direction, we assume the boost invariance, i.e., $z=\tau_0\sinh y$ and $p_z=m_T\sinh y$, where $\tau_0$,  $m_T=\sqrt{m_q^2+p_x^2+p_y^2}=\sqrt{p_x^2+p_y^2}=p_T$, and $y$ are the thermalization time, transverse mass, and rapidity, respectively.  For the initial axial charge chemical potential $\mu_5/T$, it is taken to be zero everywhere.

To study the chiral magnetic effect in heavy ion collisions, we assume as in Ref. \cite{PhysRevLett.109.202303} that the magnetic field $\mathbf{B}$ generated in these collisions is uniform along the $y$ axis, which is perpendicular to the reaction plane, and has the following time dependence:
\begin{eqnarray}
eB=\frac{eB_0}{1+(t/{\tau})^2}
\end{eqnarray}
with $B_0$ and $\tau$ being its maximum strength and lifetime, respectively.  According to Ref.~\cite{PhysRevC.85.044907}, the magnitude of the magnetic field in a heavy ion collision can be as large as several $m_{\pi}^2$ in noncentral Au+Au collisions at $\sqrt{s}=200$ GeV, although it lasts for only about 0.1 fm/$c$. However, including the electric conductivity effect and the vortical effect in QGP could increase the lifetime of the magnetic field in a heavy ion collision to several fm/$c$~\cite{McLerran2014184}.

With above initial conditions and magnetic field, we solve the anomalous transport equation using the parameters $T_0=300$ MeV, $R=3.5$ fm, $a=0.5$ fm, $c=1.5$, and $\tau_0=0.4$ fm/$c$ that are appropriate for Au+Au collisions at center-of-mass energy $\sqrt{s}=200$ GeV and impact parameter $b=9$ fm, which approximately corresponds to collisions at the 30--40\% centrality.  The total number of quarks and antiquarks is then approximately 1500 if we consider only rapidities $|y|\le 2$.  As in the AMPT model~\cite{Lin:2004en} and the Nambu-Jona-Lasinio model~\cite{Nambu:1961tp}, we assume that only quarks and antiquarks are present in the partonic phase of a heavy ion collision.  For the magnetic field, we use the values $eB_0=7 m_{\pi}^2$ and $\tau=6$ fm/$c$.  For the value of $\mu/T$, which is the same as the charge asymmetry $A_\pm=\frac{N_{+}-N_{-}}{N_{+}+N_{-}}$ of the partonic matter, where $N_+$ and $N_-$ are the numbers of positively and negatively charged quarks, respectively, we have considered a number of values between 0 and 0.16.  For the parton scattering cross section, we take it to have the temperature dependence $\sigma=\sigma_0 (T_0/T)^3$ in order to be consistent with that calculated in the NJL model~\cite{Ghosh:2015mda} and the viscosity to entropy density extracted from the measured elliptic flow using the viscous hydrodynamics~\cite{Heinz:2013th}.  We chose, however, $\sigma_0$ to reproduce the measured elliptic flow of pions.  For the temperature of the local medium where two partons scatter, it is determined from taking the local energy density to be the same as that of  an equilibrated noninteracting massless quarks and antiquarks. Because of the smaller number of partons in a local cell, the energy density is evaluated by using partons from a large ensemble of events, although only partons in the same event are allowed to scatter with each other~\cite{Bertsch:1988ik}.

Since partons in a local cell are expected to convert to hadrons when their temperature drops below the chiral restoration temperature $T_{\chi}$, which is taken to be 150 MeV, we use the duality ansatz by simply relabelling quarks as pions and letting them to follow the normal equations of motion with the velocity given by $\mathbf{p}/E$ and to scatter with known hadronic cross sections given in Ref.~\cite{Li:1995pra}.  We could import these quarks and antiquarks to the AMPT model to convert them to hadrons and let the latter undergo further hadronic scattering as in Ref.~\cite{Xu:2013sta}. Since most quarks and antiquarks are converted to resonances such as the $\rho$ meson, which subsequently decay to pions, using the quark-hadron duality by converting a quark to a pion is thus a reasonable approximation.  In the present study, we therefore evolve the partonic matter until they freeze-out kinetically, which is defined locally when the temperature of a cell drops to $T_f=120$ MeV.  We note that the effect from switching from the chiral kinetic equation to the normal equation of motion at $T_\chi$ is similar to the freeze-out hole effect discussed in Ref.~\cite{PhysRevC.89.044909}.

\section{Results}

In the present section, we study the time evolution of the partonic matter by following the motions of quarks and antiquarks according to the chiral kinetic equations [Eqs. (3) and (4)] with and without the Lorentz force as well as with and without including the chirality changing scattering between quark and antiquark.  These different studies allow us to investigate the relative importance among the effects due to the chiral kinetic motion (CKM), the Lorentz force (LF), and the chirality changing scattering (CCS).

\subsection{Pion elliptic flow}

\begin{figure}[h]
\centering
\includegraphics[width=0.5\textwidth]{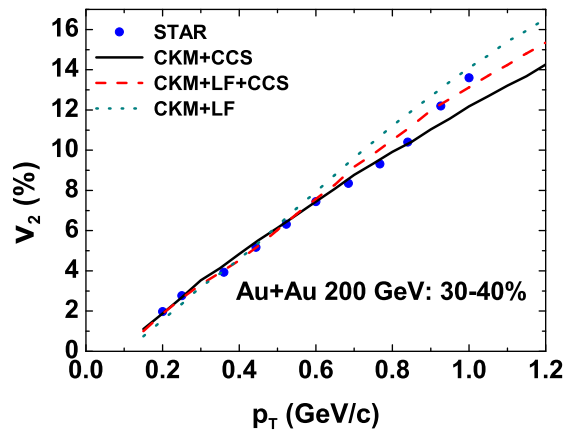}
\caption{(Color online) Elliptic flow of all kinetically freeze-out pions as a function of their transverse momentum for the three cases of chiral kinetic motion with Lorentz force and chirality changing scattering (CKM+LF+CCS), without chirality changing scattering (CKM+LF), and without Lorentz force (CKM+CCS).  Experimental data (solid circles) are from Ref.~\cite{PhysRevC.72.014904}.}
\label{17}
\end{figure}

In Fig.~\ref{17}, we first show by the dashed line the elliptic flow $v_2$ of all kinetically freeze-out pions as a function of their transverse momentum in midrapidity ($|y|\le 1$) from solving the chiral kinetic equation with Lorentz force and including the chirality changing scattering (CKM+LF+CCS). It is obtained with a coefficient $\sigma_0=13.7$ mb in the parton scattering cross section and is seen to agree with the experiment data (solid circles) from the STAR Collaboration~\cite{PhysRevC.72.014904}.  The result obtained without the chirality changing scattering (CKM+LF), using $\sigma_0=13.1$ mb and shown by the dotted line as well as that obtained without the Lorenz force (CKM+CCS), using $\sigma_0=15.5$ mb and shown by the solid line, also agree reasonably with the experimental data, particularly for momentum below 0.5 GeV.  The integrated $v_2$ of these pions with transverse momenta in the range $0.15\le p_T\le 0.5$ GeV/$c$ is 0.036 in all three cases, which reproduces surprisingly well the experimental value.

\subsection{Time evolution of eccentricity difference}

\begin{figure}[h]
\centering
\includegraphics[width=0.5\textwidth]{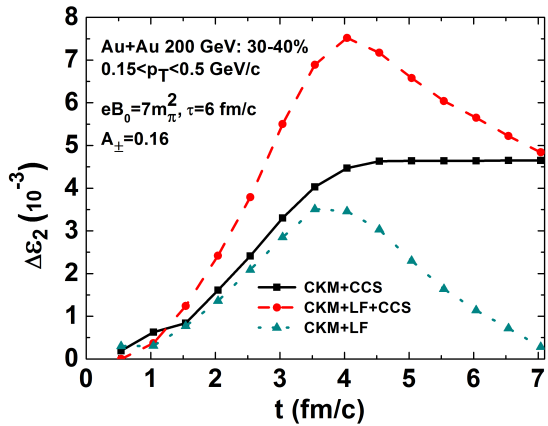}
\caption{(Color online) Eccentricity difference between positively and negatively charged particles as a function of time for different scenarios of parton dynamics as in Fig.~\ref{17} when the total charge asymmetry of the quark matter is $A_{\pm}=0.16$.}
\label{18}
\end{figure}

Figure~\ref{18} shows the time evolution of the difference $\Delta \epsilon_2=\epsilon_{2-}-\epsilon_{2+}$ between the eccentricities ($\epsilon_2=\langle (x^2-y^2)/(x^2+y^2)\rangle$) of negatively and positively charged particles for the total charge asymmetry $A_{\pm}=0.16$ of these particles. We note that this value of $A_\pm$ is much larger than that in the experimental measurements, and we use it in order to amplify the effect of the chiral magnetic wave.  The momentum range included in the calculation is again $0.15\le p_T\le 0.5$ GeV/$c$.  The solid line show the results obtained by following the motions of quarks and antiquarks via the chiral kinetic equation without the Lorentz force but including the chirality changing scattering, i.e., CKM+CCS.  The eccentricity difference between positively and negatively charged particles in this case is seen to increase with time. This increase is due to both the chiral kinetic effect, which leads to the fluctuation of axial charge, and the effect of chirality change in parton scattering, which converts the axial charge fluctuation to the charge fluctuation. How the eccentricity difference between negatively and positively charged particles change in time in the two cases of including also the Lorentz force (CKM+LF+CCS) and including the Lorentz force but without the chirality changing scattering (CKM+LF) are shown by the dashed and dotted lines, respectively. It is seen that the eccentricity difference between negatively and positively charged particles becomes larger for the case of CKM+LF+CCS and smaller for the case of CKM+LF, and in both cases it decrease after reaching a maximum value.

\subsection{Time evolution of elliptic flow difference}

\begin{figure}[h]
\centering
\includegraphics[width=0.5\textwidth]{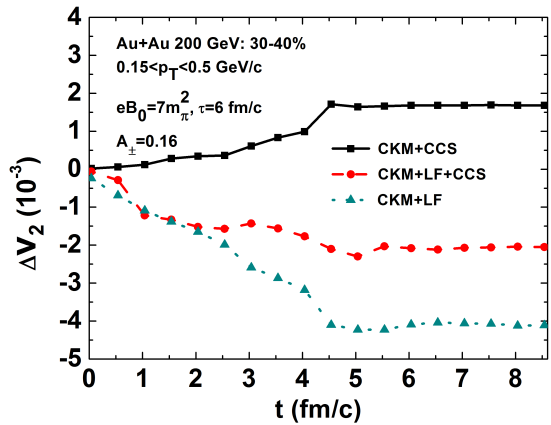}
\caption{(Color online) Same as Fig.~\ref{18} for the elliptic flow difference $\Delta v_2$ between negatively and positively charged particles.}
\label{19}
\end{figure}

\begin{figure}[h]
\centering
\includegraphics[width=0.5\textwidth]{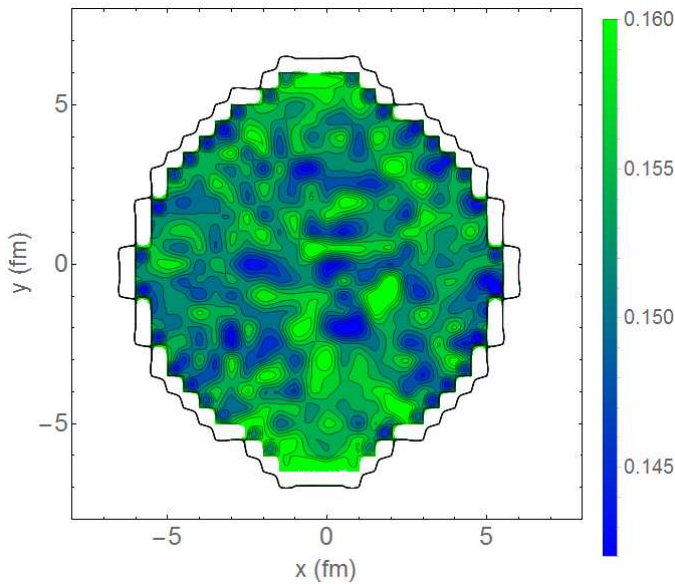}
\caption{(Color online) Charge chemical potential $\mu/T$ distribution in the transverse plane $z=0$ at time $t=5$ fm/$c$ for events with charge asymmetry $A_{\pm}=0.16$ for the case of including chiral kinetic motion and chirality changing scattering but no Lorentz force.}
\label{20}
\end{figure}

\begin{figure}[h]
\centering
\includegraphics[width=0.5\textwidth]{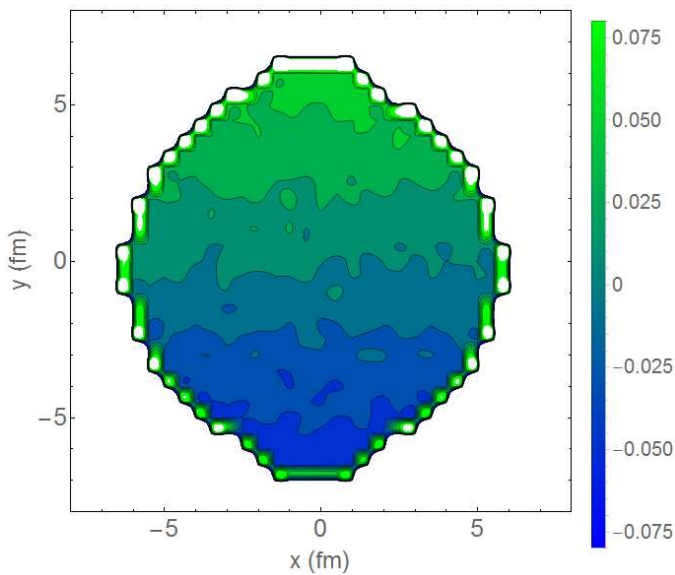}
\caption{(Color online) Same as Fig.~\ref{20} for the axial charge $\mu_5/T$.}
\label{21}
\end{figure}

Because of their different spatial eccentricities, the elliptic flows of positively and negatively charged particles become different in heavy ion collisions. In Fig.~\ref{19}, we show, for the case of CKM+CCS, by the solid line for the difference $\Delta v_2=v_{2-}-v_{2+}$ between the elliptic flows of negatively and positively charged particles with their transverse momenta between 0.15 and 0.5 GeV as a function of time for charge asymmetries $A_{\pm}=0.16$.  To better understand the reason for the elliptic flow difference between positively and negatively charged partons, we show in Figs.~\ref{20} and \ref{21} the charge chemical potential $\mu/T$ and axial charge chemical potential $\mu_5/T$ distributions of particles in the transverse plane ($z=0$) at time $t=5$ fm/$c$.

Figure \ref{20} shows that the charge chemical potential is small in the equator and large in the pole of the transverse plane of the collision, similar to that found in Ref.~\cite{PhysRevLett.107.052303} based on the CMW consideration. It also confirms the schematic pictures illustrated in Fig.~\ref{cme}. The charge quadrupole moment resulting from such a distribution would then lead to an elliptic flow that is larger for negatively charged particles than for positively charged particles as demonstrated in Ref.~\cite{Ma:2014iva} using the AMPT model by giving a finite electric quadrupole moment in the initial parton distribution.

As seen from Fig.~\ref{21}, the distribution of $\mu_5/T$ in the transverse plane of the collision shows, on the other hand, a finite dipole moment. Our results thus clearly demonstrate that the chiral kinetic equation leads to the separation of partons with different chiralities in the transverse plane of a heavy ion collision, which leads at the same time to a separation of partons of different charges by the chirality changing scattering between massless positively and negatively charged partons, reminiscent of the effect due to the CMW.   We note that if we have not allowed the change of the chiralities of these partons during their scattering in the anomalous transport model, both $\mu/T$ and $\mu_5/T$ would be uniform in space with the values $\mu/T=0.16$ and $\mu_5/T=0$.  This agrees with the results based only on the chiral kinetic equation, which is same for both positively and negatively charged partons but different for partons of different chiralities.

The results obtained after including also the Lorentz force (CKM+LF+CCS) is shown by the dashed line in Fig.~\ref{19}. It shows that the elliptic flow difference between negatively and positively charged particles now changes sign. This is because the Lorentz force has an opposite effect from that due to the chiral kinetic motion and the chirality changing quark-antiquark scattering.  This can be understood as follows.  Because of the cylindrical initial distribution, particles have a larger flow in the $z$ direction than in the $x$ direction.  With the magnetic field along the $y$ direction, the Lorentz force then deflects positively and negatively charged particles moving in the positive $z$-direction to the negative and positive $x$ axis, respectively.  In the case that the charge asymmetry $A_{\pm}$ is larger than zero, this would lead to more particles in the second and fourth quadrants than in the first and third quadrants of the $x\mbox{-}z$ plane.  Due to their more frequent collisions, positively charged particles then acquires a larger elliptic flow than the negatively charged particles. We note that if the initial particle distribution is isotropic in the $x\mbox{-}z$ plane, the charge distribution would remain isotropic in the presence of the Lorentz force, and there would be no elliptic flow difference between the positively and negatively charged particles even for $A_\pm \ne 0$.   Also shown in Fig.~\ref{19} by the dotted line are the results obtained with the inclusion of the Lorentz force but not allowing the chiralities to change in the quark-antiquark scattering (CKM+LF). It shows that this further reduces the elliptic flow difference between negatively and positively charged particles.
We note the axial charge and charge chemical potential distributions in the transverse plane for the case of CKM+LF+CCS also show finite dipole and quadrupole moments, similar to those shown in Figs. \ref{20} and \ref{21} for the case of CKM+CCS. Without the chirality changing scattering, i.e., for the case of CKM+LF, both the axial charge and charge chemical potential distributions become uniform in the transverse plane, i.e., vanishing dipole and quadrupole moments.

\subsection{Charge asymmetry dependence of the elliptic flow difference}

\begin{figure}[h]
\centering
\includegraphics[width=0.5\textwidth]{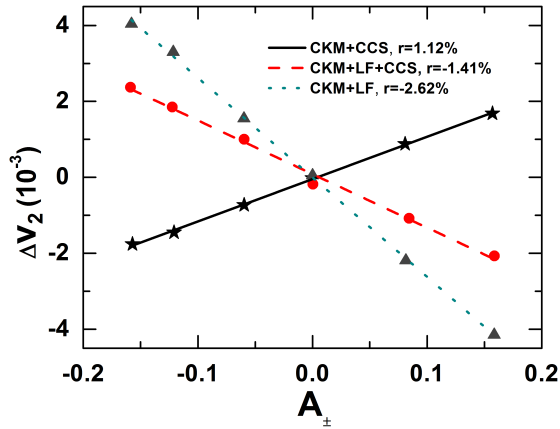}
\caption{(Color online) Elliptic flow difference $\Delta v_2$ as a function of charge asymmetry $A_{\pm}$ for different scenarios of parton dynamics as in Fig.~\ref{17}.}
\label{22}
\end{figure}

In Fig.~\ref{22}, we show by the solid line the elliptic flow difference $\Delta v_2$ between negatively and positively charged particles at freeze-out as a function of the charge asymmetry $A_{\pm}$ for the case of CKM+CCS. Again, the transverse momenta of these particles are taken to be $0.15\le p_T\le 0.5$ GeV/$c$. It is seen that the final elliptic flow difference $\Delta v_2$ is almost linearly dependent on the total charge asymmetry, similar to that found in experiments~\cite{Wang2013248c} and in the theoretical study based on the CMW~\cite{PhysRevLett.107.052303}. The slope of the $A_{\pm}$ dependence of $\Delta v_2$ in our calculation is $r=0.0112$, which gives a value $r/v_2=0.31$ that is about a factor of three smaller than the experimental value of 0.85 but is comparable to the value $r/v_2=0.27$ from a study based on the anomalous hydrodynamics~\cite{PhysRevC.89.044909}.

 For the charge asymmetry dependence of the elliptic flow difference between negatively and positively charged particles in the two cases of including also the Lorentz force (CKM+LF+CCS) and including the Lorentz force but neglecting the chirality changing quark-antiquark scattering (CKM+LF), they are shown by dashed and dotted lines, respectively. It is seen that the slope of the elliptic flow differences between negatively and positively charged particles becomes negative for these two cases with the one from CKM+LF more negative than the one from CKM+LF+CCS.

\section{Summary}

Based on the anomalous transport model, which includes the propagation of massless quarks and antiquarks according to the chiral kinetic equation and allows the change of chiralities during the scattering between positively and negatively charged partons, we have studied the elliptic flow of charged particles in non-central relativistic heavy ion collisions.  Using initial conditions from a blast wave model and assuming the presence of a strong and long-lived magnetic field, we have obtained an appreciable charge quadrupole moment in the transverse plane of the collision, which then leads to different elliptic flows for particles of negative and positive charges as the system expands. The elliptic flow difference shows a linear dependence on the total charge asymmetry of the partonic matter with a slope that is negative unless the Lorentz force is neglected. In the latter case, the result is similar to that found from studies based on the anomalous hydrodynamics, in which the Lorentz force is also neglected, using similar initial conditions and assuming similar strength and lifetime for the magnetic field. Compared to the experimental data on the elliptic flow difference between negatively and positively charged particles, ours is, however, much smaller. A larger elliptic flow difference could be obtained if we allow in the initial state a positive $\mu_5/T$ in $y>0$ and negative $\mu_5/T$ in $y<0$ besides a positive charge asymmetry, and vice versa, instead of generating such a distribution dynamically as in the present study.  Also, the different elliptic flows between charged particles could be partly due to the different mean-field potentials between particles and antiparticles in the partonic~\cite{Song:2012cd} and the hadronic~\cite{Xu:2012gf} matter of finite baryon chemical potential~\cite{Xu:2013sta}.  Nonetheless, our study does indicate that the application of the anomalous transport model based on the chiral kinetic equation to heavy ion collisions can describe the effect of the CMW on the elliptic flow of massless partons of different charges in the presence of a strong magnetic field, if quark-antiquark scatterings are dominated by the change of both their chiralities and the effect of the Lorentz force is neglected.  However, the justification for the existence of a long-lived magnetic field in relativistic heavy ion collisions remains missing, although its strength is known to be sufficiently strong~\cite{Voronyuk:2011jd,Toneev:2011aa,Toneev:2012zx}.   Also, we have not included in the present study the modification of the phase-space density in the collision term of the chiral transport model and the chiral kinetic effect resulting from the change of the momentum of massless quarks and antiquarks due to collisions~\cite{Chen:2015gta}, which is present even in the absence of magnetic field [see Eq.~(\ref{1})].  More work is needed to understand this very intriguing phenomenon that might be present in relativistic heavy ion collisions.

\section*{ACKNOWLEDGMENTS}

This work was supported in part by the US Department of Energy under Contract No. DE-SC0015266 and the Welch Foundation under Grant No. A-1358.
\bibliography{ref}

\begin{thebibliography}{47}
\expandafter\ifx\csname natexlab\endcsname\relax\def\natexlab#1{#1}\fi
\expandafter\ifx\csname bibnamefont\endcsname\relax
  \def\bibnamefont#1{#1}\fi
\expandafter\ifx\csname bibfnamefont\endcsname\relax
  \def\bibfnamefont#1{#1}\fi
\expandafter\ifx\csname citenamefont\endcsname\relax
  \def\citenamefont#1{#1}\fi
\expandafter\ifx\csname url\endcsname\relax
  \def\url#1{\texttt{#1}}\fi
\expandafter\ifx\csname urlprefix\endcsname\relax\def\urlprefix{URL }\fi
\providecommand{\bibinfo}[2]{#2}
\providecommand{\eprint}[2][]{\url{#2}}

\bibitem[{\citenamefont{Adams et~al.}(2005{\natexlab{a}})}]{Adams:2005dq}
\bibinfo{author}{\bibfnamefont{J.}~\bibnamefont{Adams}} \bibnamefont{et~al.}
  (\bibinfo{collaboration}{STAR Collaboration}), \bibinfo{journal}{Nucl. Phys.}
  \textbf{\bibinfo{volume}{A757}}, \bibinfo{pages}{102}
  (\bibinfo{year}{2005}{\natexlab{a}}).

\bibitem[{\citenamefont{Adcox et~al.}(2005)}]{Adcox:2004mh}
\bibinfo{author}{\bibfnamefont{K.}~\bibnamefont{Adcox}} \bibnamefont{et~al.}
  (\bibinfo{collaboration}{PHENIX Collaboration}), \bibinfo{journal}{Nucl.
  Phys.} \textbf{\bibinfo{volume}{A757}}, \bibinfo{pages}{184}
  (\bibinfo{year}{2005}).

\bibitem[{\citenamefont{Arsene et~al.}(2005)}]{Arsene:2004fa}
\bibinfo{author}{\bibfnamefont{I.}~\bibnamefont{Arsene}} \bibnamefont{et~al.}
  (\bibinfo{collaboration}{BRAHMS Collaboration}), \bibinfo{journal}{Nucl.
  Phys.} \textbf{\bibinfo{volume}{A757}}, \bibinfo{pages}{1}
  (\bibinfo{year}{2005}).

\bibitem[{\citenamefont{Back et~al.}(2005)}]{Back:2004je}
\bibinfo{author}{\bibfnamefont{B.~B.} \bibnamefont{Back}} \bibnamefont{et~al.},
  \bibinfo{journal}{Nucl. Phys.} \textbf{\bibinfo{volume}{A757}},
  \bibinfo{pages}{28} (\bibinfo{year}{2005}).

\bibitem[{\citenamefont{Aamodt et~al.}(2008)}]{Aamodt:2008zz}
\bibinfo{author}{\bibfnamefont{K.}~\bibnamefont{Aamodt}} \bibnamefont{et~al.}
  (\bibinfo{collaboration}{ALICE Collaboration}), \bibinfo{journal}{JINST}
  \textbf{\bibinfo{volume}{3}}, \bibinfo{pages}{S08002} (\bibinfo{year}{2008}).

\bibitem[{\citenamefont{Heinz and Snellings}(2013)}]{Heinz:2013th}
\bibinfo{author}{\bibfnamefont{U.}~\bibnamefont{Heinz}} \bibnamefont{and}
  \bibinfo{author}{\bibfnamefont{R.}~\bibnamefont{Snellings}},
  \bibinfo{journal}{Ann. Rev. Nucl. Part. Sci.} \textbf{\bibinfo{volume}{63}},
  \bibinfo{pages}{123} (\bibinfo{year}{2013}).

\bibitem[{\citenamefont{Jacobs and Wang}(2005)}]{Jacobs:2004qv}
\bibinfo{author}{\bibfnamefont{P.}~\bibnamefont{Jacobs}} \bibnamefont{and}
  \bibinfo{author}{\bibfnamefont{X.-N.} \bibnamefont{Wang}},
  \bibinfo{journal}{Prog. Part. Nucl. Phys.} \textbf{\bibinfo{volume}{54}},
  \bibinfo{pages}{443} (\bibinfo{year}{2005}).

\bibitem[{\citenamefont{Kovtun et~al.}(2005)\citenamefont{Kovtun, Son, and
  Starinets}}]{Kovtun:2004de}
\bibinfo{author}{\bibfnamefont{P.~K.} \bibnamefont{Kovtun}},
  \bibinfo{author}{\bibfnamefont{D.~T.} \bibnamefont{Son}}, \bibnamefont{and}
  \bibinfo{author}{\bibfnamefont{A.~O.} \bibnamefont{Starinets}},
  \bibinfo{journal}{Phys. Rev. Lett.} \textbf{\bibinfo{volume}{94}},
  \bibinfo{pages}{111601} (\bibinfo{year}{2005}).

\bibitem[{\citenamefont{Son and Zhitnitsky}(2004)}]{PhysRevD.70.074018}
\bibinfo{author}{\bibfnamefont{D.~T.} \bibnamefont{Son}} \bibnamefont{and}
  \bibinfo{author}{\bibfnamefont{A.~R.} \bibnamefont{Zhitnitsky}},
  \bibinfo{journal}{Phys. Rev. D} \textbf{\bibinfo{volume}{70}},
  \bibinfo{pages}{074018} (\bibinfo{year}{2004}).

\bibitem[{\citenamefont{Metlitski and Zhitnitsky}(2005)}]{PhysRevD.72.045011}
\bibinfo{author}{\bibfnamefont{M.~A.} \bibnamefont{Metlitski}}
  \bibnamefont{and} \bibinfo{author}{\bibfnamefont{A.~R.}
  \bibnamefont{Zhitnitsky}}, \bibinfo{journal}{Phys. Rev. D}
  \textbf{\bibinfo{volume}{72}}, \bibinfo{pages}{045011}
  (\bibinfo{year}{2005}).

\bibitem[{\citenamefont{Son and Sur\'owka}(2009)}]{PhysRevLett.103.191601}
\bibinfo{author}{\bibfnamefont{D.~T.} \bibnamefont{Son}} \bibnamefont{and}
  \bibinfo{author}{\bibfnamefont{P.}~\bibnamefont{Sur\'owka}},
  \bibinfo{journal}{Phys. Rev. Lett.} \textbf{\bibinfo{volume}{103}},
  \bibinfo{pages}{191601} (\bibinfo{year}{2009}).

\bibitem[{\citenamefont{Kharzeev et~al.}(2008)\citenamefont{Kharzeev, McLerran,
  and Warringa}}]{Kharzeev2008227}
\bibinfo{author}{\bibfnamefont{D.~E.} \bibnamefont{Kharzeev}},
  \bibinfo{author}{\bibfnamefont{L.~D.} \bibnamefont{McLerran}},
  \bibnamefont{and} \bibinfo{author}{\bibfnamefont{H.~J.}
  \bibnamefont{Warringa}}, \bibinfo{journal}{Nuclear Physics A}
  \textbf{\bibinfo{volume}{803}}, \bibinfo{pages}{227 } (\bibinfo{year}{2008}).

\bibitem[{\citenamefont{Fukushima et~al.}(2008)\citenamefont{Fukushima,
  Kharzeev, and Warringa}}]{PhysRevD.78.074033}
\bibinfo{author}{\bibfnamefont{K.}~\bibnamefont{Fukushima}},
  \bibinfo{author}{\bibfnamefont{D.~E.} \bibnamefont{Kharzeev}},
  \bibnamefont{and} \bibinfo{author}{\bibfnamefont{H.~J.}
  \bibnamefont{Warringa}}, \bibinfo{journal}{Phys. Rev. D}
  \textbf{\bibinfo{volume}{78}}, \bibinfo{pages}{074033}
  (\bibinfo{year}{2008}).

\bibitem[{\citenamefont{Kharzeev}(2010)}]{Kharzeev2010205}
\bibinfo{author}{\bibfnamefont{D.~E.} \bibnamefont{Kharzeev}},
  \bibinfo{journal}{Ann. Phys. (NY)} \textbf{\bibinfo{volume}{325}},
  \bibinfo{pages}{205 } (\bibinfo{year}{2010}).

\bibitem[{\citenamefont{Kharzeev and Yee}(2011)}]{PhysRevD.83.085007}
\bibinfo{author}{\bibfnamefont{D.~E.} \bibnamefont{Kharzeev}} \bibnamefont{and}
  \bibinfo{author}{\bibfnamefont{H.-U.} \bibnamefont{Yee}},
  \bibinfo{journal}{Phys. Rev. D} \textbf{\bibinfo{volume}{83}},
  \bibinfo{pages}{085007} (\bibinfo{year}{2011}).

\bibitem[{\citenamefont{Burnier et~al.}(2011)\citenamefont{Burnier, Kharzeev,
  Liao, and Yee}}]{PhysRevLett.107.052303}
\bibinfo{author}{\bibfnamefont{Y.}~\bibnamefont{Burnier}},
  \bibinfo{author}{\bibfnamefont{D.~E.} \bibnamefont{Kharzeev}},
  \bibinfo{author}{\bibfnamefont{J.}~\bibnamefont{Liao}}, \bibnamefont{and}
  \bibinfo{author}{\bibfnamefont{H.-U.} \bibnamefont{Yee}},
  \bibinfo{journal}{Phys. Rev. Lett.} \textbf{\bibinfo{volume}{107}},
  \bibinfo{pages}{052303} (\bibinfo{year}{2011}).

\bibitem[{\citenamefont{Hongo et~al.}(2013)\citenamefont{Hongo, Hirono, and
  Hirano}}]{Hongo:2013cqa}
\bibinfo{author}{\bibfnamefont{M.}~\bibnamefont{Hongo}},
  \bibinfo{author}{\bibfnamefont{Y.}~\bibnamefont{Hirono}}, \bibnamefont{and}
  \bibinfo{author}{\bibfnamefont{T.}~\bibnamefont{Hirano}}
  (\bibinfo{year}{2013}), \eprint{1309.2823}.

\bibitem[{\citenamefont{Yee and Yin}(2014)}]{PhysRevC.89.044909}
\bibinfo{author}{\bibfnamefont{H.-U.} \bibnamefont{Yee}} \bibnamefont{and}
  \bibinfo{author}{\bibfnamefont{Y.}~\bibnamefont{Yin}},
  \bibinfo{journal}{Phys. Rev. C} \textbf{\bibinfo{volume}{89}},
  \bibinfo{pages}{044909} (\bibinfo{year}{2014}).

\bibitem[{\citenamefont{Stephanov and Yin}(2012)}]{PhysRevLett.109.162001}
\bibinfo{author}{\bibfnamefont{M.~A.} \bibnamefont{Stephanov}}
  \bibnamefont{and} \bibinfo{author}{\bibfnamefont{Y.}~\bibnamefont{Yin}},
  \bibinfo{journal}{Phys. Rev. Lett.} \textbf{\bibinfo{volume}{109}},
  \bibinfo{pages}{162001} (\bibinfo{year}{2012}).

\bibitem[{\citenamefont{Son and Yamamoto}(2012)}]{Son:2012wh}
\bibinfo{author}{\bibfnamefont{D.~T.} \bibnamefont{Son}} \bibnamefont{and}
  \bibinfo{author}{\bibfnamefont{N.}~\bibnamefont{Yamamoto}},
  \bibinfo{journal}{Phys. Rev. Lett.} \textbf{\bibinfo{volume}{109}},
  \bibinfo{pages}{181602} (\bibinfo{year}{2012}).

\bibitem[{\citenamefont{Son and Yamamoto}(2013)}]{Son:2012zy}
\bibinfo{author}{\bibfnamefont{D.~T.} \bibnamefont{Son}} \bibnamefont{and}
  \bibinfo{author}{\bibfnamefont{N.}~\bibnamefont{Yamamoto}},
  \bibinfo{journal}{Phys. Rev.} \textbf{\bibinfo{volume}{D87}},
  \bibinfo{pages}{085016} (\bibinfo{year}{2013}).

\bibitem[{\citenamefont{Gao et~al.}(2012)\citenamefont{Gao, Liang, Pu, Wang,
  and Wang}}]{PhysRevLett.109.232301}
\bibinfo{author}{\bibfnamefont{J.-H.} \bibnamefont{Gao}},
  \bibinfo{author}{\bibfnamefont{Z.-T.} \bibnamefont{Liang}},
  \bibinfo{author}{\bibfnamefont{S.}~\bibnamefont{Pu}},
  \bibinfo{author}{\bibfnamefont{Q.}~\bibnamefont{Wang}}, \bibnamefont{and}
  \bibinfo{author}{\bibfnamefont{X.-N.} \bibnamefont{Wang}},
  \bibinfo{journal}{Phys. Rev. Lett.} \textbf{\bibinfo{volume}{109}},
  \bibinfo{pages}{232301} (\bibinfo{year}{2012}).

\bibitem[{\citenamefont{Chen et~al.}(2013)\citenamefont{Chen, Pu, Wang, and
  Wang}}]{PhysRevLett.110.262301}
\bibinfo{author}{\bibfnamefont{J.-W.} \bibnamefont{Chen}},
  \bibinfo{author}{\bibfnamefont{S.}~\bibnamefont{Pu}},
  \bibinfo{author}{\bibfnamefont{Q.}~\bibnamefont{Wang}}, \bibnamefont{and}
  \bibinfo{author}{\bibfnamefont{X.-N.} \bibnamefont{Wang}},
  \bibinfo{journal}{Phys. Rev. Lett.} \textbf{\bibinfo{volume}{110}},
  \bibinfo{pages}{262301} (\bibinfo{year}{2013}).

\bibitem[{\citenamefont{Manuel and Torres-Rincon}(2014)}]{Manuel:2014dza}
\bibinfo{author}{\bibfnamefont{C.}~\bibnamefont{Manuel}} \bibnamefont{and}
  \bibinfo{author}{\bibfnamefont{J.~M.} \bibnamefont{Torres-Rincon}},
  \bibinfo{journal}{Phys. Rev.} \textbf{\bibinfo{volume}{D90}},
  \bibinfo{pages}{076007} (\bibinfo{year}{2014}).

\bibitem[{\citenamefont{Chen et~al.}(2014)\citenamefont{Chen, Son, Stephanov,
  Yee, and Yin}}]{PhysRevLett.113.182302}
\bibinfo{author}{\bibfnamefont{J.-Y.} \bibnamefont{Chen}},
  \bibinfo{author}{\bibfnamefont{D.~T.} \bibnamefont{Son}},
  \bibinfo{author}{\bibfnamefont{M.~A.} \bibnamefont{Stephanov}},
  \bibinfo{author}{\bibfnamefont{H.-U.} \bibnamefont{Yee}}, \bibnamefont{and}
  \bibinfo{author}{\bibfnamefont{Y.}~\bibnamefont{Yin}},
  \bibinfo{journal}{Phys. Rev. Lett.} \textbf{\bibinfo{volume}{113}},
  \bibinfo{pages}{182302} (\bibinfo{year}{2014}).

\bibitem[{\citenamefont{Wong}(1982)}]{Wong:1982prc}
\bibinfo{author}{\bibfnamefont{C.-Y.} \bibnamefont{Wong}},
  \bibinfo{journal}{Phys. Rev. C} \textbf{\bibinfo{volume}{25}},
  \bibinfo{pages}{1460} (\bibinfo{year}{1982}).

\bibitem[{\citenamefont{Wang}(2013)}]{Wang2013248c}
\bibinfo{author}{\bibfnamefont{G.}~\bibnamefont{Wang}}, \bibinfo{journal}{Nuc.
  Phys. A} \textbf{\bibinfo{volume}{904--905}}, \bibinfo{pages}{248c }
  (\bibinfo{year}{2013}).

\bibitem[{\citenamefont{Berry}(1984)}]{Berry:1984jv}
\bibinfo{author}{\bibfnamefont{M.~V.} \bibnamefont{Berry}},
  \bibinfo{journal}{Proc. Roy. Soc. Lond.} \textbf{\bibinfo{volume}{A392}},
  \bibinfo{pages}{45} (\bibinfo{year}{1984}).

\bibitem[{\citenamefont{Satow and Yee}(2014)}]{Satow:2014lva}
\bibinfo{author}{\bibfnamefont{D.}~\bibnamefont{Satow}} \bibnamefont{and}
  \bibinfo{author}{\bibfnamefont{H.-U.} \bibnamefont{Yee}},
  \bibinfo{journal}{Phys. Rev.} \textbf{\bibinfo{volume}{D90}},
  \bibinfo{pages}{014027} (\bibinfo{year}{2014}).

\bibitem[{\citenamefont{Manuel and Torres-Rincon}(2015)}]{Manuel:2015zpa}
\bibinfo{author}{\bibfnamefont{C.}~\bibnamefont{Manuel}} \bibnamefont{and}
  \bibinfo{author}{\bibfnamefont{J.~M.} \bibnamefont{Torres-Rincon}},
  \bibinfo{journal}{Phys. Rev.} \textbf{\bibinfo{volume}{D92}},
  \bibinfo{pages}{074018} (\bibinfo{year}{2015}).

\bibitem[{\citenamefont{Ma}(2014)}]{Ma:2014iva}
\bibinfo{author}{\bibfnamefont{G.-L.} \bibnamefont{Ma}},
  \bibinfo{journal}{Phys. Lett.} \textbf{\bibinfo{volume}{B735}},
  \bibinfo{pages}{383} (\bibinfo{year}{2014}).

\bibitem[{\citenamefont{Bertsch and Das~Gupta}(1988)}]{Bertsch:1988ik}
\bibinfo{author}{\bibfnamefont{G.~F.} \bibnamefont{Bertsch}} \bibnamefont{and}
  \bibinfo{author}{\bibfnamefont{S.}~\bibnamefont{Das~Gupta}},
  \bibinfo{journal}{Phys. Rept.} \textbf{\bibinfo{volume}{160}},
  \bibinfo{pages}{189} (\bibinfo{year}{1988}).

\bibitem[{\citenamefont{Ba\ifmmode~\mbox{\c{s}}\else \c{s}\fi{}ar
  et~al.}(2012)\citenamefont{Ba\ifmmode~\mbox{\c{s}}\else \c{s}\fi{}ar,
  Kharzeev, and Skokov}}]{PhysRevLett.109.202303}
\bibinfo{author}{\bibfnamefont{G.}~\bibnamefont{Ba\ifmmode~\mbox{\c{s}}\else
  \c{s}\fi{}ar}}, \bibinfo{author}{\bibfnamefont{D.~E.}
  \bibnamefont{Kharzeev}}, \bibnamefont{and}
  \bibinfo{author}{\bibfnamefont{V.}~\bibnamefont{Skokov}},
  \bibinfo{journal}{Phys. Rev. Lett.} \textbf{\bibinfo{volume}{109}},
  \bibinfo{pages}{202303} (\bibinfo{year}{2012}).

\bibitem[{\citenamefont{Deng and Huang}(2012)}]{PhysRevC.85.044907}
\bibinfo{author}{\bibfnamefont{W.-T.} \bibnamefont{Deng}} \bibnamefont{and}
  \bibinfo{author}{\bibfnamefont{X.-G.} \bibnamefont{Huang}},
  \bibinfo{journal}{Phys. Rev. C} \textbf{\bibinfo{volume}{85}},
  \bibinfo{pages}{044907} (\bibinfo{year}{2012}).

\bibitem[{\citenamefont{McLerran and Skokov}(2014)}]{McLerran2014184}
\bibinfo{author}{\bibfnamefont{L.}~\bibnamefont{McLerran}} \bibnamefont{and}
  \bibinfo{author}{\bibfnamefont{V.}~\bibnamefont{Skokov}},
  \bibinfo{journal}{Nuc. Phys. A} \textbf{\bibinfo{volume}{929}},
  \bibinfo{pages}{184 } (\bibinfo{year}{2014}).

\bibitem[{\citenamefont{Lin et~al.}(2005)\citenamefont{Lin, Ko, Li, Zhang, and
  Pal}}]{Lin:2004en}
\bibinfo{author}{\bibfnamefont{Z.-W.} \bibnamefont{Lin}},
  \bibinfo{author}{\bibfnamefont{C.~M.} \bibnamefont{Ko}},
  \bibinfo{author}{\bibfnamefont{B.-A.} \bibnamefont{Li}},
  \bibinfo{author}{\bibfnamefont{B.}~\bibnamefont{Zhang}}, \bibnamefont{and}
  \bibinfo{author}{\bibfnamefont{S.}~\bibnamefont{Pal}},
  \bibinfo{journal}{Phys. Rev.} \textbf{\bibinfo{volume}{C72}},
  \bibinfo{pages}{064901} (\bibinfo{year}{2005}).

\bibitem[{\citenamefont{Nambu and Jona-Lasinio}(1961)}]{Nambu:1961tp}
\bibinfo{author}{\bibfnamefont{Y.}~\bibnamefont{Nambu}} \bibnamefont{and}
  \bibinfo{author}{\bibfnamefont{G.}~\bibnamefont{Jona-Lasinio}},
  \bibinfo{journal}{Phys. Rev.} \textbf{\bibinfo{volume}{122}},
  \bibinfo{pages}{345} (\bibinfo{year}{1961}).

\bibitem[{\citenamefont{Ghosh et~al.}(2016)\citenamefont{Ghosh, Peixoto, Roy,
  Serna, and Krein}}]{Ghosh:2015mda}
\bibinfo{author}{\bibfnamefont{S.}~\bibnamefont{Ghosh}},
  \bibinfo{author}{\bibfnamefont{T.~C.} \bibnamefont{Peixoto}},
  \bibinfo{author}{\bibfnamefont{V.}~\bibnamefont{Roy}},
  \bibinfo{author}{\bibfnamefont{F.~E.} \bibnamefont{Serna}}, \bibnamefont{and}
  \bibinfo{author}{\bibfnamefont{G.}~\bibnamefont{Krein}},
  \bibinfo{journal}{Phys. Rev.} \textbf{\bibinfo{volume}{C93}},
  \bibinfo{pages}{045205} (\bibinfo{year}{2016}).

\bibitem[{\citenamefont{Li and Ko}(1995)}]{Li:1995pra}
\bibinfo{author}{\bibfnamefont{B.-A.} \bibnamefont{Li}} \bibnamefont{and}
  \bibinfo{author}{\bibfnamefont{C.~M.} \bibnamefont{Ko}},
  \bibinfo{journal}{Phys. Rev.} \textbf{\bibinfo{volume}{C52}},
  \bibinfo{pages}{2037} (\bibinfo{year}{1995}).

\bibitem[{\citenamefont{Xu et~al.}(2014)\citenamefont{Xu, Song, Ko, and
  Li}}]{Xu:2013sta}
\bibinfo{author}{\bibfnamefont{J.}~\bibnamefont{Xu}},
  \bibinfo{author}{\bibfnamefont{T.}~\bibnamefont{Song}},
  \bibinfo{author}{\bibfnamefont{C.~M.} \bibnamefont{Ko}}, \bibnamefont{and}
  \bibinfo{author}{\bibfnamefont{F.}~\bibnamefont{Li}}, \bibinfo{journal}{Phys.
  Rev. Lett.} \textbf{\bibinfo{volume}{112}}, \bibinfo{pages}{012301}
  (\bibinfo{year}{2014}).

\bibitem[{\citenamefont{Adams et~al.}(2005{\natexlab{b}})\citenamefont{Adams,
  Aggarwal, Ahammed, Amonett, Anderson, Arkhipkin, Averichev, Badyal, Bai,
  Balewski et~al.}}]{PhysRevC.72.014904}
\bibinfo{author}{\bibfnamefont{J.}~\bibnamefont{Adams}},
  \bibinfo{author}{\bibfnamefont{M.~M.} \bibnamefont{Aggarwal}},
  \bibinfo{author}{\bibfnamefont{Z.}~\bibnamefont{Ahammed}},
  \bibinfo{author}{\bibfnamefont{J.}~\bibnamefont{Amonett}},
  \bibinfo{author}{\bibfnamefont{B.~D.} \bibnamefont{Anderson}},
  \bibinfo{author}{\bibfnamefont{D.}~\bibnamefont{Arkhipkin}},
  \bibinfo{author}{\bibfnamefont{G.~S.} \bibnamefont{Averichev}},
  \bibinfo{author}{\bibfnamefont{S.~K.} \bibnamefont{Badyal}},
  \bibinfo{author}{\bibfnamefont{Y.}~\bibnamefont{Bai}},
  \bibinfo{author}{\bibfnamefont{J.}~\bibnamefont{Balewski}},
  \bibnamefont{et~al.} (\bibinfo{collaboration}{STAR and STAR-RICH
  Collaborations}), \bibinfo{journal}{Phys. Rev. C}
  \textbf{\bibinfo{volume}{72}}, \bibinfo{pages}{014904}
  (\bibinfo{year}{2005}{\natexlab{b}}).

\bibitem[{\citenamefont{Ko et~al.}(2014)\citenamefont{Ko, Song, Li, Greco, and
  Plumari}}]{Song:2012cd}
\bibinfo{author}{\bibfnamefont{C.~M.} \bibnamefont{Ko}},
  \bibinfo{author}{\bibfnamefont{T.}~\bibnamefont{Song}},
  \bibinfo{author}{\bibfnamefont{F.}~\bibnamefont{Li}},
  \bibinfo{author}{\bibfnamefont{V.}~\bibnamefont{Greco}}, \bibnamefont{and}
  \bibinfo{author}{\bibfnamefont{S.}~\bibnamefont{Plumari}},
  \bibinfo{journal}{Nucl. Phys.} \textbf{\bibinfo{volume}{A928}},
  \bibinfo{pages}{234} (\bibinfo{year}{2014}).

\bibitem[{\citenamefont{Xu et~al.}(2012)\citenamefont{Xu, Chen, Ko, and
  Lin}}]{Xu:2012gf}
\bibinfo{author}{\bibfnamefont{J.}~\bibnamefont{Xu}},
  \bibinfo{author}{\bibfnamefont{L.-W.} \bibnamefont{Chen}},
  \bibinfo{author}{\bibfnamefont{C.~M.} \bibnamefont{Ko}}, \bibnamefont{and}
  \bibinfo{author}{\bibfnamefont{Z.-W.} \bibnamefont{Lin}},
  \bibinfo{journal}{Phys. Rev.} \textbf{\bibinfo{volume}{C85}},
  \bibinfo{pages}{041901(R)} (\bibinfo{year}{2012}).

\bibitem[{\citenamefont{Voronyuk et~al.}(2011)\citenamefont{Voronyuk, Toneev,
  Cassing, Bratkovskaya, Konchakovski, and Voloshin}}]{Voronyuk:2011jd}
\bibinfo{author}{\bibfnamefont{V.}~\bibnamefont{Voronyuk}},
  \bibinfo{author}{\bibfnamefont{V.~D.} \bibnamefont{Toneev}},
  \bibinfo{author}{\bibfnamefont{W.}~\bibnamefont{Cassing}},
  \bibinfo{author}{\bibfnamefont{E.~L.} \bibnamefont{Bratkovskaya}},
  \bibinfo{author}{\bibfnamefont{V.~P.} \bibnamefont{Konchakovski}},
  \bibnamefont{and} \bibinfo{author}{\bibfnamefont{S.~A.}
  \bibnamefont{Voloshin}}, \bibinfo{journal}{Phys. Rev.}
  \textbf{\bibinfo{volume}{C83}}, \bibinfo{pages}{054911}
  (\bibinfo{year}{2011}).

\bibitem[{\citenamefont{Toneev et~al.}(2012{\natexlab{a}})\citenamefont{Toneev,
  Voronyuk, Bratkovskaya, Cassing, Konchakovski, and Voloshin}}]{Toneev:2011aa}
\bibinfo{author}{\bibfnamefont{V.~D.} \bibnamefont{Toneev}},
  \bibinfo{author}{\bibfnamefont{V.}~\bibnamefont{Voronyuk}},
  \bibinfo{author}{\bibfnamefont{E.~L.} \bibnamefont{Bratkovskaya}},
  \bibinfo{author}{\bibfnamefont{W.}~\bibnamefont{Cassing}},
  \bibinfo{author}{\bibfnamefont{V.~P.} \bibnamefont{Konchakovski}},
  \bibnamefont{and} \bibinfo{author}{\bibfnamefont{S.~A.}
  \bibnamefont{Voloshin}}, \bibinfo{journal}{Phys. Rev.}
  \textbf{\bibinfo{volume}{C85}}, \bibinfo{pages}{034910}
  (\bibinfo{year}{2012}{\natexlab{a}}).

\bibitem[{\citenamefont{Toneev et~al.}(2012{\natexlab{b}})\citenamefont{Toneev,
  Konchakovski, Voronyuk, Bratkovskaya, and Cassing}}]{Toneev:2012zx}
\bibinfo{author}{\bibfnamefont{V.~D.} \bibnamefont{Toneev}},
  \bibinfo{author}{\bibfnamefont{V.~P.} \bibnamefont{Konchakovski}},
  \bibinfo{author}{\bibfnamefont{V.}~\bibnamefont{Voronyuk}},
  \bibinfo{author}{\bibfnamefont{E.~L.} \bibnamefont{Bratkovskaya}},
  \bibnamefont{and} \bibinfo{author}{\bibfnamefont{W.}~\bibnamefont{Cassing}},
  \bibinfo{journal}{Phys. Rev.} \textbf{\bibinfo{volume}{C86}},
  \bibinfo{pages}{064907} (\bibinfo{year}{2012}{\natexlab{b}}).

\bibitem[{\citenamefont{Chen et~al.}(2015)\citenamefont{Chen, Son, and
  Stephanov}}]{Chen:2015gta}
\bibinfo{author}{\bibfnamefont{J.-Y.} \bibnamefont{Chen}},
  \bibinfo{author}{\bibfnamefont{D.~T.} \bibnamefont{Son}}, \bibnamefont{and}
  \bibinfo{author}{\bibfnamefont{M.~A.} \bibnamefont{Stephanov}},
  \bibinfo{journal}{Phys. Rev. Lett.} \textbf{\bibinfo{volume}{115}},
  \bibinfo{pages}{021601} (\bibinfo{year}{2015}).

\end{thebibliography}
\end{document}